\documentclass[aps,preprint,showpacs,preprintnumbers,amsmath,amssymb]{revtex4}
\usepackage{amsmath,mathrsfs,amsbsy,color,graphicx,bm,amsthm,amsfonts}
\usepackage{units}
\usepackage{bbm}
\usepackage{times}
\usepackage{dcolumn}
\usepackage{mathrsfs}
\usepackage{amsmath,amssymb,epsfig}
\usepackage{amsmath}
\newcommand{\udots}{\mathinner{\mskip1mu\raise1pt\vbox{\kern7pt\hbox{.}}
\mskip2mu\raise4pt\hbox{.}\mskip2mu\raise7pt\hbox{.}\mskip1mu}}
\begin{document}

\title{Bosonic and fermionic mutual information of  N-partite systems in dilaton black hole background}
\author{ Xiao-Wei Teng$^1$, Rui-Yang Xu$^1$, Hui-Chen Yang$^1$,  Shu-Min Wu$^1$\footnote{Email: smwu@lnnu.edu.cn}}
\affiliation{$^1$ Department of Physics, Liaoning Normal University, Dalian 116029, China}


\begin{abstract}
We investigate multipartite  total amount of correlations by analyzing the mutual information of  N-partite states for both free bosonic and fermionic fields in the background of a Garfinkle-Horowitz-Strominger (GHS) dilaton black hole. Focusing on multipartite GHZ and W states, we examine how the Hawking effect influences the N-partite mutual information when one observer hovers near the event horizon while the remaining observers stay in the asymptotically flat region. By tracing over the inaccessible modes inside the event horizon, we derive analytical expressions for the N-partite mutual information in dilaton spacetime for both bosonic and fermionic fields. Our results show that fermionic mutual information is larger than its bosonic counterpart under the influence of the dilaton black hole,  whereas the fermionic relative entropy of coherence (REC) is smaller than the bosonic REC. Moreover, the mutual information of GHZ states is consistently larger than that of W states, while the REC of GHZ states is smaller than that of W states in curved spacetime. These findings reveal distinct behaviors of multipartite mutual information and quantum coherence for different particle statistics and multipartite state structures under gravitational effects, providing further insight into many-body  total correlations in curved spacetime.
\end{abstract}

\vspace*{0.5cm}
\maketitle
\section{Introduction}
Information is a fundamental physical quantity. From an information-theoretic perspective, the characterization and quantification of information content and correlations in physical systems are essential for understanding and optimizing information-processing tasks \cite{A1,A2,A3}. In quantum systems, correlations-encompassing both classical and quantum components play a central role in quantum information theory. Among the various correlation measures, mutual information provides a direct and comprehensive quantifier of total correlations in multipartite systems. In particular, mutual information  has proven to be a key resource in a wide range of quantum information processing tasks, including quantum communication, quantum computation, and quantum cryptography \cite{A1,A2}. It plays a pivotal role in quantifying quantum channel capacities \cite{A25,A26} and serves as an effective tool for characterizing information exchange between different representations of quantum data in quantum machine learning \cite{A27,A28}. Moreover, multipartite mutual information offers unique insights into  many-body effects beyond pairwise correlations, making it indispensable for probing phenomena such as many-body localization \cite{A29} and for quantifying quantum objectivity and the emergence of classicality in complex quantum systems \cite{A30}.

Relativistic quantum information seeks to describe and exploit quantum information tasks in regimes where relativistic effects become relevant, drawing on ideas from quantum information theory, quantum field theory, and gravitational physics \cite{SDF1,SDF2,SDF3,SDF4,SDF5,SDF6,SDF7,SDF8,SDF9,SDF10,SDF11,SDF12,SDF13,SDF14,SDF15,SDF16,SDF17,SDF18,SDF19,SDF20,SDF21,SDF22,SDF23,SDF24,SDF25,SDF27,SDF28,SDF29,SDF30,SDF31,SDF32,SDF33,SDF34,SDF35,SDF36,SDF37,SDF38,SDF39,SDF40,SDF41,SDF42,SDF43,SDF44,SDF45,SDF46,SDF47,SDF48,SDF49,SDF50,SDF51,SDF52,SDF53,SDF54,
SDF55,SDF56,AGL1,AGL2,AAA1,QFM1,QFM2,QFM3,QFM4,QFM5}. As quantum technologies continue to extend toward long-distance and high-precision implementations, the influence of spacetime curvature on quantum systems emerges as an essential factor that cannot be ignored. Gravitational environments therefore offer a natural platform for testing the robustness and redistribution of quantum resources under extreme physical conditions. Within this setting, black hole spacetimes serve as particularly instructive models for investigating the interplay between gravity and quantum correlations. Among various black hole solutions, the GHS dilaton black hole occupies a distinguished position due to its origin in the low-energy limit of string theory, where a nontrivial dilaton field modifies the spacetime geometry and alters the behavior of quantum fields \cite{J9,J10,J11}. Such modifications are expected to have a profound impact on the global structure of quantum correlations. In curved backgrounds of this kind, multipartite mutual information becomes a key quantity for assessing the performance of multiparty quantum communication, distributed quantum computation, and quantum network protocols, all of which rely on many-body correlations rather than pairwise ones. Furthermore, our analysis reveals that the response of many-body information to gravitational effects depends sensitively on both the particle statistics and the structure of the quantum states. This motivates a detailed investigation of multipartite mutual information in the GHS dilaton black hole background, providing both conceptual insight into relativistic total correlations and practical guidance for the optimal selection of quantum resources in gravitational settings.

Based on the above motivations,  in this work, we investigate N-partite  mutual information for both free bosonic and fermionic fields under the influence of the  GHS dilaton black hole.  We initially consider N-particle GHZ and W states prepared in the asymptotically flat region. Subsequently, one observer is assumed to hover near the event horizon of the black hole, while the remaining observers stay static in the asymptotically flat region. Since the field modes inside the event horizon are inaccessible, we trace over these modes and obtain analytical expressions of the N-partite mutual information for both bosonic and fermionic fields.
We find that, as the dilaton increases, the multipartite mutual information decreases monotonically and approaches a constant value, which becomes independent of the field mode frequency in the extreme black hole limit.  Our analysis reveals two notable features of  total amount of correlations  in the dilaton black hole background. First, we observe a clear distinction between fermionic and bosonic fields: under the influence of the dilaton, fermionic multipartite mutual information remains higher than its bosonic counterpart, indicating that fermionic correlations are more robust against gravitational degradation. Interestingly, the behavior of REC exhibits the opposite trend, with fermionic REC being smaller than the bosonic one, suggesting that the coherence properties of fermionic states respond differently to spacetime curvature than their total correlations. Second, the comparison between different multipartite state structures highlights the role of correlation patterns: GHZ states consistently retain higher mutual information than W states,  whereas the REC of GHZ states is lower than that of W states.  
Together, these findings demonstrate that both particle statistics and multipartite state structure play important roles in shaping the behavior of different  correlation measures in curved spacetime and provide further insight into many-body  correlations in gravitational backgrounds.

The paper is organized as follows. In Sec.II, we discuss the quantization of both free bosonic
and fermionic fields in the background of the GHS dilaton black hole. In Sec.III, we study bosonic
and fermionic mutual information of N-partite GHZ and W states in the dilaton black hole. Finally,
Sec.IV summarizes our main results.

\section{Quantization of  bosonic and fermionic fields in GHS dilaton spacetime}
The metric of a charged GHS dilaton black hole spacetime is given by
\begin{eqnarray}\label{S7}
ds^{2}=-\bigg(\frac{r-2M}{r-2\mathcal{D}}\bigg)dt^{2}+\bigg(\frac{r-2M}{r-2\mathcal{D}}\bigg)^{-1}dr^{2}+r(r-2\mathcal{D})d\Omega^{2},
\end{eqnarray}
where $\mathcal{D}$ and $M$ denote the dilaton field and the mass of the black hole, respectively \cite{J51}. These parameters are related to the magnetic charge $Q$ through $\mathcal{D}=Q^{2}/2M$. In this paper, we adopt natural units where $G=c=\hbar=\kappa_{B}=1$.

\subsection{Bosonic field}
In this dilaton spacetime,  the dynamics of a massless scalar field can be governed by the
Klein-Gordon equation
\begin{eqnarray}\label{S8}
\frac{1}{\sqrt{-g}}\partial_{\mu}(\sqrt{-g}g^{\mu\nu}\partial_{\nu})\Psi=0,
\end{eqnarray}
where  $\Psi$ is the scalar field \cite{J61,J62}.
The normal mode solution can be expressed as
\begin{eqnarray}\label{S9}
\Psi_{\omega lm}=\frac{1}{h(r)}\chi_{\omega l}(r)Y_{lm}(\theta,\varphi)e^{-i\omega t},
\end{eqnarray}
where $Y_{lm}(\theta,\varphi)$ represents a scalar spherical harmonic on the unit two-sphere and $h(r)=\sqrt{r(r-2\mathcal{D})}$. The radial equation can be written as
\begin{eqnarray}\label{S10}
\frac{d^{2}\chi_{\omega l}}{dr^{2}_{\ast}}+[\omega^{2}-V(r)]\chi_{\omega l}=0,
\end{eqnarray}
and
\begin{eqnarray}\label{S11}
V(r)=\frac{f(r)}{h(r)}\frac{d}{dr}\bigg[f(r)\frac{dh(r)}{dr}\bigg]+\frac{f(r)l(l+1)}{h^{2}(r)}+f(r)\bigg[\mu^{2}+\frac{2\xi \mathcal{D}^{2}(r-2M)}{r^{2}(r-2\mathcal{D})^{3}}\bigg],
\end{eqnarray}
where $f(r)=(r-2M)/(r-2\mathcal{D})$ and the tortoise coordinate $r_{\ast}$ is defined  as $dr_{\ast}=dr/f(r)$ \cite{J51}.

Solving Eq.(\ref{S10}) near the event horizon, one obtains an incoming wave function that is analytic everywhere in the spacetime manifold

\begin{eqnarray}\label{S12}
\Psi_{\mathrm{in},\omega lm}=e^{-i\omega v}Y_{lm}(\theta,\varphi),
\end{eqnarray}
and the outgoing wave functions for the regions outside and inside the event horizon are given by
\begin{eqnarray}\label{S13}
\Psi_{\mathrm{out},\omega lm}(r>r_{+})=e^{-i\omega u}Y_{lm}(\theta,\varphi),
\end{eqnarray}
\begin{eqnarray}\label{S14}
\Psi_{\mathrm{out},\omega lm}(r<r_{+})=e^{i\omega u}Y_{lm}(\theta,\varphi),
\end{eqnarray}
where $u=t-r_{\ast}$ and $v=t+r_{\ast}$. The solutions given in Eqs.(\ref{S13}) and (\ref{S14}) respectively possess analyticity in the exterior and interior regions of the event horizon, and thus span a complete orthogonal family. The generalized light-like Kruskal coordinates are subsequently defined as
\begin{eqnarray}\label{S15}
u&=&-4(M-\mathcal{D})\ln[-U/(4M-4\mathcal{D})],\notag\\v&=&4(M-\mathcal{D})\ln[V/(4M-4\mathcal{D})],\mathrm{if}\;r>r_{+},\notag\\
u&=&-4(M-\mathcal{D})\ln[U/(4M-4\mathcal{D})],\notag\\v&=&4(M-\mathcal{D})\ln[V/(4M-4\mathcal{D})],\mathrm{if}\;r<r_{+}.
\end{eqnarray}
Therefore, Eqs.(\ref{S13}) and (\ref{S14}) can be rephrased as
\begin{eqnarray}\label{S16}
\Phi_{\mathrm{out},\omega lm}(r>r_{+})=e^{4(M-\mathcal{D})i\omega\ln[U/(4M-4\mathcal{D})]}Y_{lm}(\theta,\varphi),
\end{eqnarray}
\begin{eqnarray}\label{S17}
\Phi_{\mathrm{out},\omega lm}(r<r_{+})=e^{-4(M-\mathcal{D})i\omega\ln[-U/(4M-4\mathcal{D})]}Y_{lm}(\theta,\varphi).
\end{eqnarray}
By utilizing the relation $-1=e^{i\pi}$ and ensuring that  Eq.(\ref{S16}) is analytic in the lower half-plane of $U$, a complete basis for positive-energy $U$ modes can be written as
\begin{eqnarray}\label{S18}
\Phi_{\mathrm{I},\omega lm}=e^{2\pi\omega(M-\mathcal{D})}\Phi_{\mathrm{out},\omega lm}(r>r_{+})
+e^{-2\pi\omega(M-\mathcal{D})}\Phi_{\mathrm{out},\omega lm}^{\ast}(r<r_{+}),
\end{eqnarray}
\begin{eqnarray}\label{S19}
\Phi_{\mathrm{II},\omega lm}=e^{-2\pi\omega(M-\mathcal{D})}\Phi_{\mathrm{out},\omega lm}^{\ast}(r>r_{+})
+e^{2\pi\omega(M-\mathcal{D})}\Phi_{\mathrm{out},\omega lm}(r<r_{+}).
\end{eqnarray}
Since Eqs.(\ref{S18}) and (\ref{S19}) are analytic over the entire real range of $U$ and $V$, they define a complete basis of positive-frequency modes. Consequently, the functions $\Phi_{\mathrm{I},\omega lm}$ and $\Phi_{\mathrm{II},\omega lm}$ serve as the appropriate mode functions for quantizing the field in Kruskal spacetime \cite{J88}.

By applying second quantization on the field in the exterior region of the dilaton black hole, thereby obtaining the Bogoliubov transformations for the creation and annihilation operators in both dilaton and Kruskal spacetimes
\begin{eqnarray}\label{S20}
a_{K,\omega lm}^{B,\dagger}=\frac{1}{\sqrt{1-e^{-8\pi\omega(M-\mathcal{D})}}}b_{\mathrm{out},\omega lm}^{B,\dag}
-\frac{1}{\sqrt{e^{8\pi\omega(M-\mathcal{D})}-1}}b_{\mathrm{in},\omega lm}^{B},
\end{eqnarray}
\begin{eqnarray}\label{S21}
a_{K,\omega lm}^{B}=\frac{1}{\sqrt{1-e^{-8\pi\omega(M-\mathcal{D})}}}b_{\mathrm{out},\omega lm}^{B}-\frac{1}{\sqrt{e^{8\pi\omega(M-\mathcal{D})}-1}}b_{\mathrm{in},\omega lm}^{B,\dag},
\end{eqnarray}
where  $a_{K,\omega lm}^{B,\dagger}$ and $a_{K,\omega lm}^{B}$ denote the bosonic creation and annihilation operators associated with the global Kruskal modes, while $b_{\mathrm{in},\omega lm}^{B,\dagger}$, $b_{\mathrm{in},\omega lm}^{B}$, and $b_{\mathrm{out},\omega lm}^{B,\dagger}$, $b_{\mathrm{out},\omega lm}^{B}$ correspond to the field modes defined inside and outside the event horizon of the dilaton black hole, respectively. Therefore, the Kruskal vacuum $|0\rangle_{K}^{B}$ can be defined as $a_{K,\omega lm}^{B}|0\rangle_{K}^{B}=0$.
After proper normalization, the Kruskal vacuum of the bosonic field in dilaton spacetime is a maximally entangled two-mode squeezed state
\begin{eqnarray}\label{S23}
|0\rangle_{K}^{B}=\sqrt{1-e^{-8\pi\omega(M-\mathcal{D})}}\sum^{\infty}_{p=0} e^{-4p\pi\omega(M-\mathcal{D})}|p\rangle_{\mathrm{out}}^{B}|p\rangle_{\mathrm{in}}^{B},
\end{eqnarray}
and the first excited state of the bosonic field reads
\begin{eqnarray}\label{S24}
|1\rangle_{K}^{B}=a_{K,\omega lm}^{B,\dagger}|0\rangle _{K}^{B}=
\big[1-e^{-8\pi\omega(M-\mathcal{D})}\big]\sum^{\infty}_{p=0}\sqrt{p+1} e^{-4p\pi\omega(M-\mathcal{D})}|p+1\rangle_{\mathrm{out}}^{B}|p\rangle_{\mathrm{in}}^{B},
\end{eqnarray}
where $B$ represents the bosonic field, $\{|p\rangle_{\mathrm{out}}\}$ and $\{|p\rangle_{\mathrm{in}}\}$ are the orthonormal bases for the outside and inside regions of the event horizon, respectively \cite{J98}.
For an external observer situated outside the dilaton black hole, it is imperative to trace over the modes within the interior region. This necessity arises because these modes are situated in a causally disconnected region, rendering them inaccessible to the observer. As a result, the Hawking radiation spectrum for the bosonic field is derived as
\begin{eqnarray}\label{S25}
N^{B}_{\omega}=\frac{1}{e^{8\pi\omega(M-\mathcal{D})}-1}.
\end{eqnarray}
Eq.(\ref{S25}) reveals that an observer positioned outside the GHS dilaton black hole perceives a thermal Bose-Einstein distribution of particles while traversing the Kruskal vacuum.

\subsection{Fermionic field}
In analogy with the bosonic field, the fermionic creation and annihilation operators in the Kruskal and dilaton frames are related by the Bogoliubov transformations
\begin{eqnarray}\label{S26}
a^{F,\dagger}_{K,\omega lm}=\frac{1}{\sqrt{e^{-8\pi\omega(M-\mathcal{D})}+1}}a^{F,\dagger}_{\mathrm{out},\omega lm}-\frac{1}{\sqrt{e^{8\pi\omega(M-\mathcal{D})}+1}}b_{\mathrm{in},\omega lm}^{F},
\end{eqnarray}
\begin{eqnarray}\label{S27}
a^{F}_{K,\omega lm}=\frac{1}{\sqrt{e^{-8\pi\omega(M-\mathcal{D})}+1}}a_{\mathrm{out},\omega lm}^{F}-\frac{1}{\sqrt{e^{8\pi\omega(M-\mathcal{D})}+1}}b^{F,\dagger}_{\mathrm{in},\omega lm},
\end{eqnarray}
where the superscript $F$ represents the fermionic field. Consequently, the Kruskal vacuum and the first excited state in dilaton spacetime can be expressed as
\begin{eqnarray}\label{S28}
|0\rangle_{K}^{F}=\frac{1}{\sqrt{e^{-8\pi\omega(M-\mathcal{D})}+1}}|0\rangle_{\mathrm{out}}^{F}|0\rangle_{\mathrm{in}}^{F}+\frac{1}{\sqrt{e^{8\pi\omega(M-\mathcal{D})}+1}}|1\rangle_{\mathrm{out}}^{F}|1\rangle_{\mathrm{in}}^{F},
\end{eqnarray}
and
\begin{eqnarray}\label{S29}
|1\rangle_{K}^{F}=|1\rangle_{\mathrm{out}}^{F}|0\rangle_{\mathrm{in}}^{F},
\end{eqnarray}
with $\{|p\rangle_{\mathrm{out}}^{F}\}$ and $\{|p\rangle_{\mathrm{in}}^{F}\}$ denoting the orthonormal bases for the exterior and interior fermionic modes, respectively \cite{C2,C4,C5,C6}.
The Hawking radiation spectrum for the fermionic field can be derived as
\begin{eqnarray}\label{S30}
N^{F}_{\omega}=\frac{1}{e^{8\pi\omega(M-\mathcal{D})}+1},
\end{eqnarray}
indicating that an observer outside the event horizon detects a thermal Fermi-Dirac distribution of particles.
By comparing Eqs.(\ref{S25}) and (\ref{S30}), it becomes evident that the Bose-Einstein and Fermi-Dirac statistics lead to fundamentally different gravitational responses in dilaton spacetime.
This statistical distinction is reflected in the contrasting behaviors of N-partite mutual information 
for bosonic and fermionic fields, highlighting the decisive role of quantum statistics in governing the near-horizon dynamics of each system around a dilaton black hole.

\section{N-partite mutual information of bosonic and fermionic fields in dilaton spacetime}
We consider a system of $N$ observers $(N\geq3)$, all initially stationed at the same point in the asymptotically flat region of the dilaton black hole. Observer $O_{i}$ is associated with mode $i$ $(i=1,2,\ldots,N)$. The shared quantum state among them is either a GHZ state
\begin{eqnarray}\label{S31}
|\mathrm{GHZ}_{123\ldots N}^{B/F}\rangle=\frac{1}{\sqrt{2}}\left(|0_{1}0_{2}\ldots0_{N-1}0_{N}\rangle+|1_{1}1_{2}\ldots1_{N-1}1_{N}\rangle\right),
\end{eqnarray}
or a W state
\begin{eqnarray}\label{S32}
|\mathrm{W}_{123\ldots N}^{B/F}\rangle=\frac{1}{\sqrt{N}}\left(|1_{1}0_{2}\ldots0_{N-1}0_{N}\rangle+|0_{1}1_{2}\ldots0_{N-1}0_{N}\rangle+\ldots+|0_{1}0_{2}\ldots0_{N-1}1_{N}\rangle\right),
\end{eqnarray}
where the superscripts $B$ and $F$ denote bosonic and fermionic modes, respectively. After the qubits are distributed, one observer (the $N$-th) moves to hover near the event horizon of the dilaton black hole, while the remaining $N-1$ observers remain stationary in the asymptotically flat region.

\subsection{Bosonic mutual information of  N-partite systems}
Using the transformations between Kruskal and dilaton coordinates given in Eqs.(\ref{S23}) and (\ref{S24}), the N-partite bosonic GHZ state in Eq.(\ref{S31}) can be expressed in terms of exterior and interior modes as
\begin{eqnarray}\label{S33}
|\mathrm{GHZ}^{B}_{123\ldots N+1}\rangle &=&\frac{1}{\sqrt{2}}\bigg\{[1-e^{-8\pi\omega(M-\mathcal{D})}]^{\frac{1}{2}}\overbrace{(|0\rangle_{1}|0\rangle_{2}\cdots|0\rangle_{N-1})}^{|\overline{0}\rangle}\sum^{\infty}_{p=0}\big[e^{-4p\pi\omega(M-\mathcal{D})}\notag\\
&&|p\rangle_{N_{\mathrm{out}}}|p\rangle_{N+1_{\mathrm{in}}}\big]+[1-e^{-8\pi\omega(M-\mathcal{D})}]\overbrace{(|1\rangle_{1}|1\rangle_{2}\cdots|1\rangle_{N-1})}^{|\overline{1}\rangle}\notag\\
&&\sum^{\infty}_{q=0}\big[e^{-4q\pi\omega(M-\mathcal{D})}\sqrt{q+1}|q+1\rangle_{N_{\mathrm{out}}}|q\rangle_{N+1_{\mathrm{in}}}\big]\bigg\},
\end{eqnarray}
where $|\overline{0}\rangle=|0\rangle_{1}|0\rangle_{2}\cdots|0\rangle_{N-1}$ and $|\overline{1}\rangle=|1\rangle_{1}|1\rangle_{2}\cdots|1\rangle_{N-1}$ represent the collective states of the first $N-1$ observers.
As the region outside the event horizon is causally disconnected from the interior, the modes inside the horizon are inaccessible. Tracing over these interior modes yields the mixed density matrix for the exterior region
\begin{eqnarray}\label{S34}
\rho_{123\ldots N_{\mathrm{out}}}^{B,\mathrm{GHZ}}
&=&\frac{1}{2}\sum_{p=0}^{\infty}e^{-8p\pi\omega(M-\mathcal{D})}\bigg\{[1-e^{-8\pi\omega(M-\mathcal{D})}]|\overline{0}\rangle\langle\overline{0}||p\rangle_{N_{\mathrm{out}}}\langle p|\notag\\
&+&[1-e^{-8\pi\omega(M-\mathcal{D})}]^{\frac{3}{2}}\sqrt{p+1}|\overline{0}\rangle\langle\overline{1}||p\rangle_{N_{\mathrm{out}}}\langle p+1|\notag\\
&+&[1-e^{-8\pi\omega(M-\mathcal{D})}]^{\frac{3}{2}}\sqrt{p+1}|\overline{1}\rangle\langle\overline{0}|| p+1\rangle_{N_{\mathrm{out}}}\langle p|\notag\\
&+&[1-e^{-8\pi\omega(M-\mathcal{D})}]^{2}(p+1)|\overline{1}\rangle\langle\overline{1}|| p+1\rangle_{N_{\mathrm{out}}}\langle p+1|\bigg\}.
\end{eqnarray}
The resulting mixed state can be expressed in the density-matrix form, which exhibits a block-diagonal structure composed of $2\times2$ blocks along the main diagonal, while all off-diagonal elements vanish. Explicitly, it can be written as
\begin{eqnarray}\label{S35}
\rho_{123\ldots N_{\mathrm{out}}}^{B,\mathrm{GHZ}}=\frac{1}{2}\begin{pmatrix}
0 & & & & & \\
& \delta_0 & & & & \\
& & \delta_1 & & & \\
& & & \ddots & & \\
& & & & \delta_p & \\
& & & & & \ddots \\
\end{pmatrix},
\end{eqnarray}
where the $p$-th block $(p=0,1,2,\ldots)$ is given by
\begin{eqnarray}\label{S36}
\delta_{p}(\rho_{123\ldots N_{\mathrm{out}}}^{B,\mathrm{GHZ}})=\begin{pmatrix}
(1-e^{-8\pi\omega(M-\mathcal{D})})e^{-8p\pi \omega(M-\mathcal{D})} & \sqrt{p+1} (1-e^{-8\pi\omega(M-\mathcal{D})})^{\frac{3}{2}}e^{-8p\pi \omega(M-\mathcal{D})} \\
\sqrt{p+1} (1-e^{-8\pi\omega(M-\mathcal{D})})^{\frac{3}{2}}e^{-8p\pi\omega(M-\mathcal{D})} & (p+1) (1-e^{-8\pi\omega(M-\mathcal{D})})^{2}e^{-8p\pi\omega(M-\mathcal{D})}  \\
\end{pmatrix}.
\end{eqnarray}
The eigenvalues of this $p$-th block of the state $\rho_{123\ldots N_{\mathrm{out}}}^{B,\mathrm{GHZ}}$ are 0 and
\begin{eqnarray}\label{S37}
\eta_{123\ldots N_{\mathrm{out}}}^{B,\mathrm{GHZ}}=\frac{1}{2}[1-e^{-8\pi\omega(M-\mathcal{D})}]e^{-8p\pi\omega(M-\mathcal{D})}[2+p-(p+1)e^{-8\pi\omega(M-\mathcal{D})}].
\end{eqnarray}
The trace of the density matrix is therefore $\mathrm{Tr}[\rho_{123\ldots N_{\mathrm{out}}}^{B,\mathrm{GHZ}}]= \displaystyle \sum_{p=0}^{\infty}\eta_{123\ldots N_{\mathrm{out}}}^{B,\mathrm{GHZ}}=1$.

We first outline the notion of mutual information as a measure of total correlations in multipartite quantum systems \cite{C8}.  For a composite system composed of two subsystems $A$ and $B$,  the total amount of correlations shared between them is quantified by the mutual information, defined as
\begin{eqnarray}\label{a1}
I(\rho_{AB})=S(\rho_{A})+S(\rho_{B})-S(\rho_{AB}),
\end{eqnarray}
where $\rho_{A}=\mathrm{Tr_{B}}[\rho_{AB}]$ and $\rho_{B}=\mathrm{Tr_{A}}[\rho_{AB}]$  denote the reduced density matrices of subsystems $A$ and $B$, respectively. 
Here  $S(\rho)=-\mathrm{Tr}[\rho\mathrm{log}_{2}(\rho)]$  is the von Neumann entropy, which characterizes the information content of a quantum state. Although the bipartite mutual information captures all correlations between two subsystems, many realistic quantum information protocols involve more than two parties. 
To fully characterize correlations in such scenarios, it is necessary to extend the concept of mutual information to multipartite systems. 
For an N-partite quantum state described by the density matrix $\rho_{123\ldots N}$, the multipartite mutual information is defined as
\begin{eqnarray}\label{a2}
I(\rho_{123\ldots N})=\sum_{i=1}^{N}S(\rho_i)-S(\rho_{123\ldots N}),
\end{eqnarray}
where $\rho_i$ denotes the reduced density matrix of the $i$-th subsystem.
In the physical setting considered in this work, where one observer is located near the black hole horizon while the remaining observers stay in the asymptotically flat region, the above expression can be recast into a more convenient form.  After tracing out the inaccessible interior modes, the multipartite mutual information becomes
\begin{eqnarray}\label{a3}
I(\rho_{123\ldots N})=(N-1)S(\rho_{1/2/3/\ldots/N-1})+S(\rho_{N_{\mathrm{out}}})-S(\rho_{123\ldots N_{\mathrm{out}}}),
\end{eqnarray}
where  $\rho_{N_{\mathrm{out}}}$ denotes the reduced density matrix of the exterior mode associated with the observer near the event horizon, and $\rho_{123\ldots N_{\mathrm{out}}}$  represents the global density matrix obtained after tracing over the interior (inaccessible) degrees of freedom.

To evaluate the multipartite mutual information, it is necessary to first compute the von Neumann entropy of each individual subsystem. The reduced density matrix for the exterior mode of the last qubit, denoted as $\rho_{N_{\mathrm{out}}}^{B,\mathrm{GHZ}}$, is obtained by taking the partial trace over the first $N-1$ particles of the total exterior state: $\rho_{N_{\mathrm{out}}}^{B,\mathrm{GHZ}}=\mathrm{Tr}_{123\ldots N-1}[\rho_{123\ldots N_{\mathrm{out}}}^{B,\mathrm{GHZ}}]$, and is given by
\begin{eqnarray}\label{S38}
\rho_{N_{\mathrm{out}}}^{B,\mathrm{GHZ}}&=&\frac{1}{2}\big[1-e^{-8\pi\omega(M-\mathcal{D})}\big]\sum_{p=0}^{\infty}e^{-8p\pi\omega(M-\mathcal{D})}\bigg\{1+p\notag\\
&\times&\big[1-e^{-8\pi\omega(M-\mathcal{D})}\big]e^{8\pi\omega(M-\mathcal{D})}\bigg\} |p\rangle_{N_{\mathrm{out}}}\langle p|.
\end{eqnarray}
This is an infinite-dimensional diagonal matrix, where the $p$-th eigenvalue is given by
\begin{eqnarray}\label{S39}
\eta_{N_{\mathrm{out}}}^{B,\mathrm{GHZ}}=\frac{1}{2}[1-e^{-8\pi\omega(M-\mathcal{D})}]e^{-8p\pi\omega(M-\mathcal{D})}\bigg\{1+p[1-e^{-8\pi\omega(M-\mathcal{D})}]e^{8\pi\omega(M-\mathcal{D})}\bigg\},
\end{eqnarray}
which satisfy the normalization condition $\mathrm{Tr}[\rho_{N_{\mathrm{out}}}^{B,\mathrm{GHZ}}]= \displaystyle \sum_{p=0}^{\infty}\eta_{N_{\mathrm{out}}}^{B,\mathrm{GHZ}}=1$.
The reduced density matrix $\rho_{1/2/\cdots/N-1}^{B,\mathrm{GHZ}}$ for the first $N-1$ observers is obtained by tracing out all other particles, yielding
\begin{eqnarray}\label{S40}
\rho_{1/2/\cdots/N-1}^{B,\mathrm{GHZ}}=\frac{1}{2}\big(|1\rangle\langle1|+|0\rangle\langle0|\big).
\end{eqnarray}
According to the definition of von Neumann entropy $S(\rho) = -\mathrm{Tr}(\rho\log_2\rho)$, we can calculate the entropies of $\rho_{1/2/\cdots/N-1}^{B,\mathrm{GHZ}}$, $\rho_{N_{\mathrm{out}}}^{B,\mathrm{GHZ}}$ and $\rho_{123\ldots N_{\mathrm{out}}}^{B,\mathrm{GHZ}}$, respectively. 
Substituting them into Eq.(\ref{a3}), the mutual information for the bosonic GHZ state $\rho_{123\ldots N_{\mathrm{out}}}^{B,\mathrm{GHZ}}$ in dilaton spacetime is obtained as
\begin{eqnarray}\label{a4}
I^{\mathrm{GHZ}}_{B}(N)&=&(N-1)-\frac{1}{2}[1-e^{-8\pi\omega(M-\mathcal{D})}]\sum_{p=0}^{\infty}e^{-8p\pi\omega(M-\mathcal{D})}\big(1+p[1-e^{-8\pi\omega(M-\mathcal{D})}]\notag\\
&\times&e^{8\pi\omega(M-\mathcal{D})}\big)\log_{2}\bigg\{\frac{1}{2}[1-e^{-8\pi\omega(M-\mathcal{D})}]e^{-8p\pi\omega(M-\mathcal{D})}\big(1+p[1-e^{-8\pi\omega(M-\mathcal{D})}]\notag\\
&\times&e^{8\pi\omega(M-\mathcal{D})}\big)\bigg\}+\frac{1}{2}[1-e^{-8\pi\omega(M-\mathcal{D})}]\sum_{p=0}^{\infty}e^{-8p\pi\omega(M-\mathcal{D})}\big(2+p-(p+1)\notag\\
&\times&e^{-8\pi\omega(M-\mathcal{D})}\big)\log_{2}\bigg\{\frac{1}{2}[1-e^{-8\pi\omega(M-\mathcal{D})}]e^{-8p\pi\omega(M-\mathcal{D})}\big(2+p-(p+1)\notag\\
&\times&e^{-8\pi\omega(M-\mathcal{D})}\big)\bigg\}.
\end{eqnarray}

\begin{figure}
\begin{minipage}[t]{0.5\linewidth}
\centering
\includegraphics[width=3.0in,height=5.2cm]{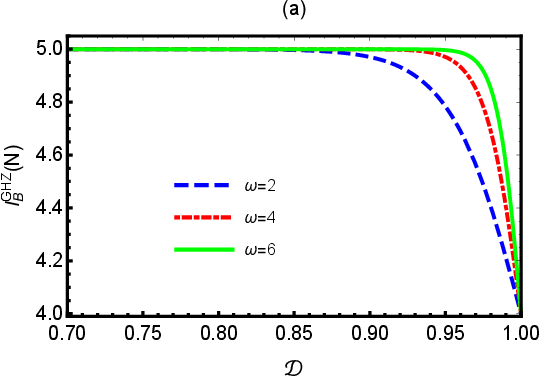}
\label{fig1a}
\end{minipage}%
\begin{minipage}[t]{0.5\linewidth}
\centering
\includegraphics[width=3.0in,height=5.2cm]{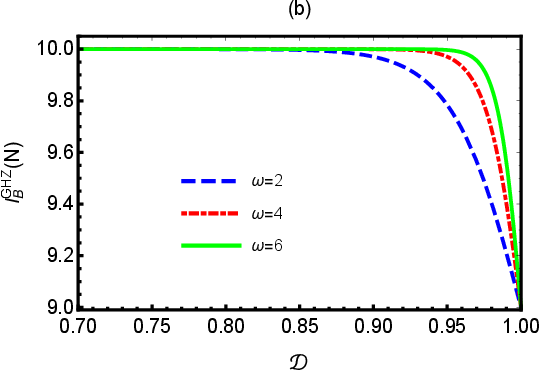}
\label{fig1b}
\end{minipage}%
\caption{Multipartite mutual information $I^{\mathrm{GHZ}}_{B}(N)$ of bosonic GHZ state as a function of the dilaton $\mathcal{D}$ for systems with $N=5$ (a) and $N=10$ (b), shown at different field mode frequencies $\omega$.}
\label{Fig.1}
\end{figure}

In Fig.\ref{Fig.1}, we plot the behavior of the mutual information $I^{\mathrm{GHZ}}_{B}(N)$ of GHZ state for the bosonic field as a function of the dilaton  $\mathcal{D}$ under different mode frequencies $\omega$. Fig.\ref{Fig.1}(a) and (b) correspond to systems with particle numbers $N=5$ and $N=10$, respectively. It can be observed that the mutual information is consistently larger for $N=10$ than for $N=5$, indicating that states involving more particles contain 
the total amount of correlations and are less susceptible to dilaton-induced information loss. From a physical perspective, for GHZ state, increasing the number of particles enhances the  mutual information of the system in curved spacetime; however, the amount of quantum entanglement remains unchanged with increasing particle number \cite{C9}. In both cases, the mutual information remains nearly constant for small values of $\mathcal{D}$ and then decreases sharply as $\mathcal{D}\rightarrow M$ (corresponding to the case of an extreme black hole), reflecting the accelerating degradation of mutual information under stronger dilaton effects. Furthermore, higher frequencies $\omega$ correspond to larger mutual information, suggesting that high-frequency modes are more resistant to dilaton-induced decoherence. However, in the extreme limit, the mutual information saturates to a value that is independent of the mode frequency. These findings highlight the fundamental role of particle numbers $N$ and frequency $\omega$ in preserving  mutual information in curved spacetime, offering valuable insights for future relativistic quantum information processing.

Following the same procedure as for the GHZ state, the pure N-partite W state can be expanded in terms of dilaton modes as
\begin{eqnarray}\label{S41}
|\mathrm{W}^{B,\mathrm{W}}_{123\ldots N+1}\rangle
&=&\frac{1}{\sqrt{N}}\bigg\{[1-e^{-8\pi\omega(M-\mathcal{D})}]^{\frac{1}{2}}|1\rangle_{_{1}}|0\rangle_{_{2}}\cdots|0\rangle_{_{N-1}}\sum^{\infty}_{p=0}\big[e^{-4p\pi\omega(M-\mathcal{D})}|p\rangle_{N_{\mathrm{out}}}|p\rangle_{N+1_{\mathrm{in}}}\big]\notag\\
&+&[1-e^{-8\pi\omega(M-\mathcal{D})}]^{\frac{1}{2}}|0\rangle_{_{1}}|1\rangle_{_{2}}\cdots|0\rangle_{_{N-1}}\sum^{\infty}_{p=0}\big[e^{-4p\pi\omega(M-\mathcal{D})}|p\rangle_{N_{\mathrm{out}}}|p\rangle_{N+1_{\mathrm{in}}}\big]\notag\\
&+&\cdots+[1-e^{-8\pi\omega(M-\mathcal{D})}]^{\frac{1}{2}}|0\rangle_{_{1}}|0\rangle_{_{2}}\cdots|1\rangle_{_{N-1}}\sum^{\infty}_{p=0} \big[ e^{-4p\pi\omega(M-\mathcal{D})}|p\rangle_{N_{\mathrm{out}}}|p\rangle_{N+1_{\mathrm{in}}}\big]\notag\\
&+&[1-e^{-8\pi\omega(M-\mathcal{D})}]|\overline{0}\rangle\sum^{\infty}_{p=0}\big[\sqrt{p+1}e^{-4p\pi\omega(M-\mathcal{D})}|p+1\rangle_{N_{\mathrm{out}}}|p\rangle_{N+1_{\mathrm{in}}}\big]\bigg\}.
\end{eqnarray}
After tracing over the inaccessible interior modes, the reduced density operator for the exterior region becomes
\begin{eqnarray}\label{S42}
\rho_{123\ldots N_{\mathrm{out}}}^{B,\mathrm{W}}=\frac{1}{N}\big(\rho_{\mathrm{diag}}^{B,\mathrm{W}}+\rho_{\mathrm{off}}^{B,\mathrm{W}}\big),
\end{eqnarray}
with the diagonal part
\[
\begin{aligned}
\rho_{\mathrm{diag}}^{B,\mathrm{W}} &= \sum_{i=1}^{N-1} |1\rangle_{i} \langle 1| \bigotimes_{j=1, j \neq i}^{N-1} |0\rangle_{j} \langle 0| [1 - e^{-8\pi \omega (M - \mathcal{D})}] \sum_{p=0}^{\infty} e^{-8p\pi \omega (M - \mathcal{D})} |p\rangle_{N_{\mathrm{out}}} \langle p| \\
&\quad + \bigotimes_{i=1}^{N-1} |0\rangle_{i} \langle 0| [1 - e^{-8\pi \omega (M - \mathcal{D})}]^2 \sum_{p=0}^{\infty} (p+1) e^{-8p\pi \omega (M - \mathcal{D})} |p+1\rangle_{N_{ \mathrm{out}}} \langle p+1|,
\end{aligned}
\]
and the off‑diagonal part
\[
\begin{aligned}
\rho_{\mathrm{off}}^{B,\mathrm{W}} &=\sum_{i,j=1(i\neq j)}^{N-1}|0\rangle_{i}\langle1||1\rangle_{j}\langle0|\bigotimes_{k=1(k\neq i,k\neq j)}^{N-1}|0\rangle_{k}\langle0|[1-e^{-8\pi\omega(M-\mathcal{D})}]\sum_{p=0}^{\infty}e^{-8p\pi\omega(M-\mathcal{D})}|p\rangle_{N_{\mathrm{out}}}\langle p| \\
&+\sum_{i=1}^{N-1}|1\rangle_{i}\langle0|\bigotimes_{j=1(j\neq i)}^{N-1}|0\rangle_{j}\langle0|[1-e^{-8\pi\omega(M-\mathcal{D})}]^{\frac{3}{2}}\sum_{p=0}^{\infty}\sqrt{p+1}e^{-8p\pi\omega(M-\mathcal{D})}|p\rangle_{N_{\mathrm{out}}}\langle p+1| \\
&+\sum_{i=1}^{N-1}|0\rangle_{i}\langle1|\bigotimes_{j=1(j\neq i)}^{N-1}|0\rangle_{j}\langle0|[1-e^{-8\pi\omega(M-\mathcal{D})}]^{\frac{3}{2}}\sum_{p=0}^{\infty}\sqrt{p+1}e^{-8p\pi\omega(M-\mathcal{D})}|p+1\rangle_{N_{\mathrm{out}}}\langle p|.
\end{aligned}
\]
The reduced density matrix $\rho_{123\ldots N_{\mathrm{out}}}^{B,\mathrm{W}}$ possesses a block-diagonal structure and can be expressed as
\begin{eqnarray}\label{S43}
\rho_{123\ldots N_{\mathrm{out}}}^{B,\mathrm{W}}=\frac{1}{N}\begin{pmatrix}
0 & & & & & \\
& \delta'_0 & & & & \\
& & \delta'_1 & & & \\
& & & \ddots & & \\
& & & & \delta'_p & \\
& & & & & \ddots \\
\end{pmatrix}.
\end{eqnarray}
In the ordered basis $\{|10\ldots0p\rangle, |01\ldots0p\rangle, \cdots, |00\ldots1p\rangle, |00\ldots0,p+1\rangle \} $, each diagonal block $\delta'_p$ takes the form
\begin{eqnarray}\label{S44}
&&\delta'_{p}(\rho_{123\ldots N_{\mathrm{out}}}^{B,W})=\notag\\
&&\beta^{p}\begin{pmatrix}
(1-\beta) & (1-\beta) & \cdots & (1-\beta) & (1-\beta)^\frac{3}{2}\sqrt{p+1}\\
(1-\beta) & (1-\beta) & \cdots & (1-\beta) & (1-\beta)^\frac{3}{2}\sqrt{p+1}\\
\vdots & \vdots & \ddots &\vdots & \vdots \\
(1-\beta) & (1-\beta) & \cdots & (1-\beta) & (1-\beta)^\frac{3}{2}\sqrt{p+1}\\
(1-\beta)^\frac{3}{2}\sqrt{p+1} & (1-\beta)^\frac{3}{2}\sqrt{p+1} & \cdots & (1-\beta)^\frac{3}{2}\sqrt{p+1} & (1-\beta)^{2}(p+1)
\end{pmatrix},
\end{eqnarray}
where $\beta=e^{-8\pi\omega(M-\mathcal{D})}$. The matrix $\delta'_{p}(\rho_{123\ldots N_{\mathrm{out}}}^{B,\mathrm{W}})$ is of size $N\times N$. Its upper-left $(N-1)\times(N-1)$ submatrix consists entirely of the same element $(1-\beta)\beta^p$. The entries in both the last row and the last column are all equal to $(1-\beta)^{\frac{3}{2}}\sqrt{p+1}\beta^{p}$, with the exception of the bottom-right corner element at position $(N,N)$, which is $(1-\beta)^{2}(p+1)\beta^p$.

The eigenvalues of this $p$-th block of the density matrix $\rho_{123\ldots N_{\mathrm{out}}}^{B,\mathrm{W}}$ are 0 (with multiplicity $N-1$) and one nonzero eigenvalue
\begin{eqnarray}\label{S45}
\lambda^{B,\mathrm{W}}_{123\ldots N_{\mathrm{out}}}=\frac{1}{N}e^{-8p\pi\omega(M-\mathcal{D})}[1-e^{-8\pi\omega(M-\mathcal{D})}][N+p-(p+1)e^{-8\pi\omega(M-\mathcal{D})}].
\end{eqnarray}
The reduced density matrix of the exterior mode for the observer near the horizon is obtained by tracing over the first $N-1$ subsystems
\begin{eqnarray}\label{S46}
\rho_{N_{\mathrm{out}}}^{B,\mathrm{W}} &=& \frac{1}{N}\sum_{p=0}^{\infty}e^{-8p\pi\omega(M-\mathcal{D})}[1-e^{-8\pi\omega(M-\mathcal{D})}]\bigg\{[1-e^{-8\pi\omega(M-\mathcal{D})}](p+1)|p+1\rangle_{N_{\mathrm{out}}}\langle p+1|\notag\\
&+& (N-1)|p\rangle_{N_{\mathrm{out}}}\langle p|\bigg\}=\frac{1}{N}e^{-8p\pi\omega(M-\mathcal{D})}[1-e^{-8\pi\omega(M-\mathcal{D})}]\bigg\{N-1+pe^{8\pi\omega(M-\mathcal{D})}\notag\\
&\times&[1-e^{-8\pi\omega(M-\mathcal{D})}]\bigg\}|p\rangle_{N_{\mathrm{out}}}\langle p|.
\end{eqnarray}
This operator is diagonal in the particle‑number basis, and its $p$-th eigenvalue is
\begin{eqnarray}\label{S47}
\lambda^{B,\mathrm{W}}_{N_{\mathrm{out}}}=\frac{1}{N}e^{-8p\pi\omega(M-\mathcal{D})}[1-e^{-8\pi\omega(M-\mathcal{D})}]\bigg\{N-1+pe^{8\pi\omega(M-\mathcal{D})}[1-e^{-8\pi\omega(M-\mathcal{D})}]\bigg\}.
\end{eqnarray}
The reduced density matrix for any one of the first $N-1$ observers follows from tracing over all other parties:
\begin{eqnarray}\label{a5}
\rho_{1/2/\cdots/N-1}^{B,\mathrm{W}}=\frac{1}{N}\big(|1\rangle\langle1|+(N-1)|0\rangle\langle0|\big).
\end{eqnarray}
Inserting the von Neumann entropies of $\rho_{1/2/\cdots/N-1}^{B,\mathrm{W}}$, $\rho_{N_{\mathrm{out}}}^{B,\mathrm{W}}$, and $\rho_{123\ldots N_{\mathrm{out}}}^{B,\mathrm{W}}$ into the mutual information formula Eq.(\ref{a3}) yields
\begin{eqnarray}\label{S48}
I^{\mathrm{W}}_{B}(N)&=&-(N-1)(\frac{1}{N}\log_{2}\frac{1}{N}+\frac{N-1}{N}\log_{2}\frac{N-1}{N})-\frac{1}{N}\sum_{p=0}^{\infty}e^{-8p\pi\omega(M-\mathcal{D})}\notag\\
&\times&[1-e^{-8\pi\omega(M-\mathcal{D})}]\big(N-1+pe^{8\pi\omega(M-\mathcal{D})}[1-e^{-8\pi\omega(M-\mathcal{D})}]\big)\log_{2}\bigg\{\frac{1}{N}\notag\\
&\times&e^{-8p\pi\omega(M-\mathcal{D})}[1-e^{-8\pi\omega(M-\mathcal{D})}]\big\{N-1+pe^{8\pi\omega(M-\mathcal{D})}[1-e^{-8\pi\omega(M-\mathcal{D})}]\big\}\bigg\}\notag\\
&+&\frac{1}{N}\sum_{p=0}^{\infty}e^{-8p\pi\omega(M-\mathcal{D})}[1-e^{-8\pi\omega(M-\mathcal{D})}][N+p-(p+1)e^{-8\pi\omega(M-\mathcal{D})}]\log_{2}\bigg\{\frac{1}{N}\notag\\
&\times&e^{-8p\pi\omega(M-\mathcal{D})}[1-e^{-8\pi\omega(M-\mathcal{D})}][N+p-(p+1)e^{-8\pi\omega(M-\mathcal{D})}]\bigg\}.
\end{eqnarray}

\begin{figure}
\begin{minipage}[t]{0.5\linewidth}
\centering
\includegraphics[width=3.0in,height=5.2cm]{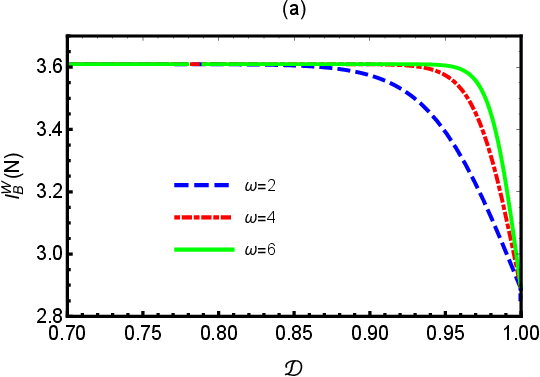}
\label{fig2a}
\end{minipage}%
\begin{minipage}[t]{0.5\linewidth}
\centering
\includegraphics[width=3.0in,height=5.2cm]{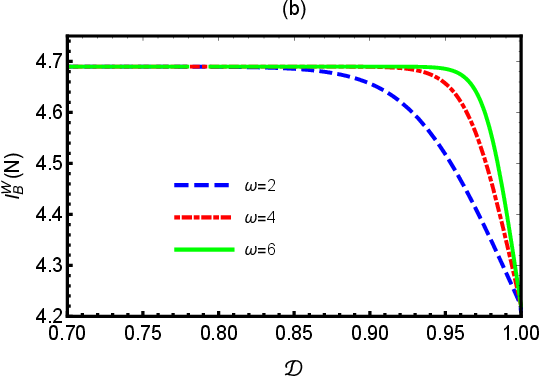}
\label{fig2b}
\end{minipage}%
\caption{Mutual information $I^{\mathrm{W}}_{B}(N)$ for W state of bosonic field as a function of the dilaton $\mathcal{D}$ for systems with $N=5$ (a) and $N=10$ (b), shown at different field mode frequencies $\omega$.}
\label{Fig.2}
\end{figure}

In Fig.\ref{Fig.2}, we present the mutual information $I^{\mathrm{W}}_{B}(N)$ for W state of bosonic field, calculated from Eq.(\ref{S48}), as a function of the dilaton $\mathcal{D}$.
The results are shown for systems with $N=5$ and $N=10$ at various field mode frequencies $\omega$. 
Similar to the GHZ case, the mutual information for W state exhibits a monotonic decline with increasing $\mathcal{D}$, indicating that the dilaton background universally degrades 
the total amount of correlations through mode mixing and information loss into inaccessible regions. 
From  Fig.\ref{Fig.2}, we find that the multipartite mutual information of W state is systematically lower than that of GHZ state in the dilaton spacetime.  Moreover, increasing the initial particle number leads to an enhancement of the multipartite mutual information of W state in the background of the dilaton black hole, while their quantum entanglement gradually decreases \cite{C10,C11}. 
This behavior highlights that, for W state, multipartite mutual information and multipartite entanglement respond differently to gravitational effects and play distinct roles in relativistic quantum information tasks. Although the multipartite entanglement of W state is progressively suppressed in the dilaton black hole background, their multipartite mutual information can still increase with the number of particles, indicating that total correlations remain an effective resource even when entanglement is degraded.

\subsection{Fermionic mutual information of  N-partite systems}
We follow a procedure similar to that applied to the bosonic GHZ state: using the dilaton modes from
Eqs.(\ref{S28}) and (\ref{S29}), we reformulate Eq.(\ref{S31}) for the fermionic field as
\begin{eqnarray}\label{S49}
&&|\mathrm{GHZ}^{F}_{123\ldots N+1}\rangle=\frac{1}{\sqrt{2}}\bigg\{\overbrace{\big(|0\rangle_{1}|0\rangle_{2}\cdots|0\rangle_{N-1}\big)}^{|\bar{0}\rangle}\{[e^{-8\pi\omega(M-\mathcal{D})}+1]^{-\frac{1}{2}}|0\rangle_{N_{\mathrm{out}}}|0\rangle_{N+1_{\mathrm{in}}}\notag\\
&&+[e^{8\pi\omega(M-\mathcal{D})}+1]^{-\frac{1}{2}}|1\rangle_{N_{\mathrm{out}}}|1\rangle_{N+1_{\mathrm{in}}}\}+\overbrace{\big(|1\rangle_{1}|1\rangle_{2}\cdots|1\rangle_{N-1}\big)}^{|\bar{1}\rangle}|1\rangle_{N_{\mathrm{out}}}|0\rangle_{N+1_{\mathrm{in}}}\bigg\}.
\end{eqnarray}
By performing a partial trace over the inaccessible modes inside the event horizon,  the reduced density operator $\rho_{123\ldots N_{\mathrm{out}}}^{F,\mathrm{GHZ}}$ can be obtained as
\begin{eqnarray}\label{S50}
\rho^{F,\mathrm{GHZ}}_{123\ldots N_{\mathrm{out}}}&=&\frac{1}{2}\bigg\{|\overline{0}\rangle\langle\overline{0}|\big\{[e^{-8\pi\omega(M-\mathcal{D})}+1]^{-1}|0\rangle_{N_{\mathrm{out}}}\langle0|+[e^{8\pi\omega(M-\mathcal{D})}+1]^{-1}|1\rangle_{N_{\mathrm{out}}}\langle1|\big\}\notag\\
&+&|\overline{0}\rangle\langle\overline{1}|[e^{-8\pi\omega(M-\mathcal{D})}+1]^{-\frac{1}{2}}|0\rangle_{N_{\mathrm{out}}}\langle1|+|\overline{1}\rangle\langle\overline{0}|[e^{-8\pi\omega(M-\mathcal{D})}+1]^{-\frac{1}{2}}|1\rangle_{N_{\mathrm{out}}}\langle0|\notag\\
&+&|\overline{1}\rangle\langle\overline{1}||1\rangle_{N_{\mathrm{out}}}\langle1|\bigg\}.
\end{eqnarray}
with its  eigenvalues 
\begin{eqnarray}\label{S51}
\eta^{F,\mathrm{GHZ}}_{123\ldots N_{\mathrm{out}}}=\frac{1}{2[1+e^{8\pi\omega(M-\mathcal{D})}]}, \quad \eta^{'F,\mathrm{GHZ}}_{123\ldots N_{\mathrm{out}}}=\frac{1+2e^{8\pi\omega(M-\mathcal{D})}}{2[1+e^{8\pi\omega(M-\mathcal{D})}]}.
\end{eqnarray}
By tracing over the degrees of freedom associated with the first $N-1$ particles, we derive the reduced density matrix  $\rho_{N_{\mathrm{out}}}^{F,\mathrm{GHZ}}$,  which takes the form 
\begin{eqnarray}\label{S52}
\rho_{N_{\mathrm{out}}}^{F,\mathrm{GHZ}}=\frac{1}{2}\bigg\{[1+e^{-8\pi\omega(M-\mathcal{D})}]^{-1}|0\rangle_{N_{\mathrm{out}}}\langle0|+\big\{[e^{8\pi\omega(M-\mathcal{D})}+1]^{-1}+1\big\}|1\rangle_{N_{\mathrm{out}}}\langle1|\bigg\},
\end{eqnarray}
with eigenvalues:
\begin{eqnarray}\label{S53}
\eta^{F,\mathrm{GHZ}}_{N_{\mathrm{out}}}=\frac{e^{8\pi\omega(M-\mathcal{D})}}{2[1+e^{8\pi\omega(M-\mathcal{D})}]}, \quad \eta^{'F,\mathrm{GHZ}}_{N_{\mathrm{out}}}=\frac{2+e^{8\pi\omega(M-\mathcal{D})}}{2[1+e^{8\pi\omega(M-\mathcal{D})}]}.
\end{eqnarray}
Tracing out subsystems of the rest particles, we obtain reduced density matrix $\rho_{1/2/\cdots/N-1}^{F,\mathrm{GHZ}}$ as
\begin{eqnarray}\label{a6}
\rho_{1/2/\cdots/N-1}^{F,\mathrm{GHZ}}=\frac{1}{2}\big(|1\rangle\langle1|+|0\rangle\langle0|\big).
\end{eqnarray}
Using the von Neumann entropy, we evaluate the entropies of density $\rho_{1/2/\cdots/N-1}^{F,\mathrm{GHZ}}$, $\rho_{N_{\mathrm{out}}}^{F,\mathrm{GHZ}}$ and $\rho_{123\ldots N_{\mathrm{out}}}^{F,\mathrm{GHZ}}$, respectively. Inserting these results into Eq.(\ref{a3}) gives the mutual information of the state $\rho^{F,\mathrm{GHZ}}_{123\ldots N_{\mathrm{out}}}$ in dilaton spacetime
\begin{eqnarray}\label{S54}
I^{\mathrm{GHZ}}_{F}(N)&=&(N-1)-\frac{e^{8\pi\omega(M-\mathcal{D})}}{2[1+e^{8\pi\omega(M-\mathcal{D})}]}\log_{2}\bigg\{\frac{e^{8\pi\omega(M-\mathcal{D})}}{2[1+e^{8\pi\omega(M-\mathcal{D})}]}\bigg\}-\frac{2+e^{8\pi\omega(M-\mathcal{D})}}{2[1+e^{8\pi\omega(M-\mathcal{D})}]}\notag\\
&\times&\log_{2}\bigg\{\frac{2+e^{8\pi\omega(M-\mathcal{D})}}{2[1+e^{8\pi\omega(M-\mathcal{D})}]}\bigg\}+\frac{1}{2[1+e^{8\pi\omega(M-\mathcal{D})}]}\log_{2}\bigg\{\frac{1}{2[1+e^{8\pi\omega(M-\mathcal{D})}]}\bigg\}\notag\\
&+&\frac{1+2e^{8\pi\omega(M-\mathcal{D})}}{2[1+e^{8\pi\omega(M-\mathcal{D})}]}\log_{2}\bigg\{\frac{1+2e^{8\pi\omega(M-\mathcal{D})}}{2[1+e^{8\pi\omega(M-\mathcal{D})}]}\bigg\}.
\end{eqnarray}

\begin{figure}
\begin{minipage}[t]{0.5\linewidth}
\centering
\includegraphics[width=3.0in,height=5.2cm]{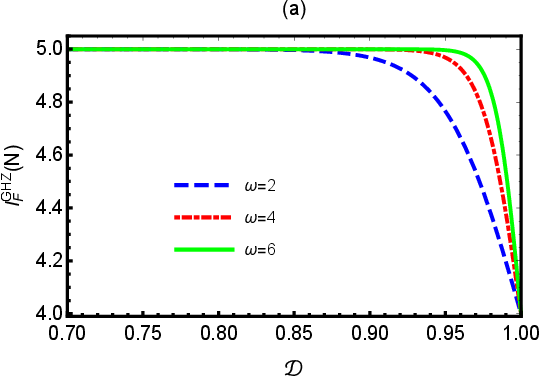}
\label{fig3a}
\end{minipage}%
\begin{minipage}[t]{0.5\linewidth}
\centering
\includegraphics[width=3.0in,height=5.2cm]{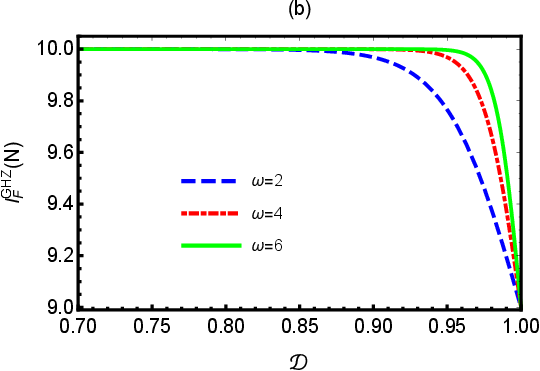}
\label{fig3b}
\end{minipage}%
\caption{Multipartite mutual information $I^{\mathrm{GHZ}}_{F}(N)$ of fermionic GHZ state as a function of the dilaton  $\mathcal{D}$, shown for systems with $N=5$ and $N=10$ particles at different field mode frequencies $\omega$.}
\label{Fig.3}
\end{figure}

In Fig.\ref{Fig.3}, we plot the mutual information $I^{\mathrm{GHZ}}_{F}(N)$ for GHZ state of the fermionic field as a function of the dilaton  $\mathcal{D}$ comparing systems with $N=5$ and $N=10$ particles at various mode frequencies $\omega$, derived from Eq.(\ref{S54}).  It is observed that, in the asymptotically flat region of the dilaton black hole, the mutual information for both bosonic and fermionic fields coincides at the same initial value. As the dilaton  $\mathcal{D}$ increases, the mutual information of fermionic fields remains consistently higher than that of bosonic fields, reflecting the stronger robustness of fermionic correlations against gravitational degradation.

The pure N-partite W state of the fermionic field [see Eq.(\ref{S32})] in the asymptotically flat region can be expressed in terms of the dilaton modes defined in Eqs.(\ref{S28}) and (\ref{S29}) as
\begin{eqnarray}\label{S55}
|\mathrm{W}^{F}_{123\ldots N+1}\rangle
&=&\frac{1}{\sqrt{N}}\bigg\{|1\rangle_{_{1}}|0\rangle_{_{2}}\cdots|0\rangle_{_{N-1}}\{[e^{-8\pi\omega(M-\mathcal{D})}+1]^{-\frac{1}{2}}|0\rangle_{N_{\mathrm{out}}}|0\rangle_{N+1_{\mathrm{in}}}\notag\\
&+&[e^{8\pi\omega(M-\mathcal{D})}+1]^{-\frac{1}{2}}|1\rangle_{N_{\mathrm{out}}}|1\rangle_{N+1_{\mathrm{in}}}\}+|0\rangle_{_{1}}|1\rangle_{_{2}}\cdots|0\rangle_{_{N-1}}\big\{[e^{-8\pi\omega(M-\mathcal{D})}+1]^{-\frac{1}{2}}\notag\\
&\times&|0\rangle_{N_{\mathrm{out}}}|0\rangle_{N+1_{\mathrm{in}}}+[e^{8\pi\omega(M-\mathcal{D})}+1]^{-\frac{1}{2}}|1\rangle_{N_{\mathrm{out}}}|1\rangle_{N+1_{\mathrm{in}}}\big\}+\cdots\notag\\
&+&|0\rangle_{_{1}}|0\rangle_{_{2}}\cdots|1\rangle_{_{N-1}}\big\{[e^{-8\pi\omega(M-\mathcal{D})}+1]^{-\frac{1}{2}}|0\rangle_{N_{\mathrm{out}}}|0\rangle_{N+1_{\mathrm{in}}}+[e^{8\pi\omega(M-\mathcal{D})}+1]^{-\frac{1}{2}}\notag\\
&\times&|1\rangle_{N_{\mathrm{out}}}|1\rangle_{N+1_{\mathrm{in}}}\big\}+|0\rangle_{1}|0\rangle_{2}\cdots|0\rangle_{N-1}|1\rangle_{N_{\mathrm{out}}}|0\rangle_{N+1_{\mathrm{in}}}\bigg\}.
\end{eqnarray}
By tracing over the field modes inaccessible beyond the event horizon, the reduced density operator for the fermionic  N-partite W state outside the horizon $\rho_{123\ldots N_{\mathrm{out}}}^{F,\mathrm{W}}$ can be expressed as
\begin{eqnarray}\label{S56}
\rho_{123\ldots N_{\mathrm{out}}}^{F,\mathrm{W}}=\frac{1}{N}\big(\rho_{\mathrm{diag}}^{F}+\rho_{\mathrm{off}}^{F}\big),
\end{eqnarray}
where the diagonal part $\rho_{\mathrm{diag}}^{F}$ reads
\[
\begin{aligned}
\rho_{\mathrm{diag}}^{F}&=\sum_{i=1}^{N-1}|1\rangle_{i}\langle1|\bigotimes_{j=1(j\neq i)}^{N-1}|0\rangle_{j}\langle0|\big\{[e^{-8\pi\omega(M-\mathcal{D})}+1]^{-1}|0\rangle_{N_{\mathrm{out}}}\langle0| \\
&+[e^{8\pi\omega(M-\mathcal{D})}+1]^{-1}|1\rangle_{N_{\mathrm{out}}}\langle1|\big\}+\bigotimes_{i=1}^{N-1}|0\rangle_{i}\langle0||1\rangle_{N_{\mathrm{out}}}\langle1|,
\end{aligned}
\]
and the off-diagonal part $\rho_{\mathrm{off}}^{F}$ becomes 
\[
\begin{aligned}
\rho_{\mathrm{off}}^{F}&=\sum_{i,j=1(i\neq j)}^{N-1}|1\rangle_{i}\langle0||0\rangle_{j}\langle1|\bigotimes_{k=1,(k\neq i,k\neq j)}^{N-1}|0\rangle_{k}\langle0|\bigg\{[e^{-8\pi\omega(M-\mathcal{D})}+1]^{-1}|0\rangle_{N_{\mathrm{out}}}\langle0| \\
&+[e^{8\pi\omega(M-\mathcal{D})}+1]^{-1}|1\rangle_{N_{\mathrm{out}}}\langle1|\bigg\}+\sum_{i=1}^{N-1}|1\rangle_{i}\langle0|\bigotimes_{j=1(j\neq i)}^{N-1}|0\rangle_{j}\langle0|[e^{-8\pi\omega(M-\mathcal{D})}+1]^{-\frac{1}{2}}|0\rangle_{N_{\mathrm{out}}}\langle1|\\
&+\sum_{i=1}^{N-1}|0\rangle_{i}\langle1|\bigotimes_{j=1(j\neq i)}^{N-1}|0\rangle_{j}\langle0|[e^{-8\pi\omega(M-\mathcal{D})}+1]^{-\frac{1}{2}}|1\rangle_{N_{\mathrm{out}}}\langle0|.
\end{aligned}
\]
After solving for the eigenvalues of this density operator, we obtain the following eigenvalues as
\begin{eqnarray}\label{S57}
\lambda^{F,\mathrm{W}}_{123\ldots N_{\mathrm{out}}}=\frac{N-1}{N[1+e^{8\pi\omega(M-\mathcal{D})}]}, \quad \lambda^{'F,\mathrm{W}}_{123\ldots N_{\mathrm{out}}}=\frac{1+Ne^{8\pi\omega(M-\mathcal{D})}}{N[1+e^{8\pi\omega(M-\mathcal{D})}]}.
\end{eqnarray}
The reduced density matrix of a single particle outside the horizon, $\rho_{N_{\mathrm{out}}}^{F,\mathrm{W}}$, 
is obtained by tracing over the remaining  $N-1$ subsystems
\begin{eqnarray}\label{S58}
\rho_{N_{\mathrm{out}}}^{F,\mathrm{W}} &=&\frac{1}{N}\bigg\{(N-1)[e^{-8\pi\omega(M-\mathcal{D})}+1]^{-1}|0\rangle_{N_{\mathrm{out}}}\langle0|\notag\\
&+&\{(N-1)[e^{8\pi\omega(M-\mathcal{D})}+1]^{-1}+1\}|1\rangle_{N_{\mathrm{out}}}\langle1|\bigg\}.
\end{eqnarray}
The eigenvalues of this reduced density matrix are then
\begin{eqnarray}\label{S59}
\lambda^{F,\mathrm{W}}_{N_{\mathrm{out}}}=\frac{(N-1)e^{8\pi\omega(M-\mathcal{D})}}{N[1+e^{8\pi\omega(M-\mathcal{D})}]}, \quad \lambda^{'F,\mathrm{W}}_{N_{\mathrm{out}}}=\frac{N+e^{8\pi\omega(M-\mathcal{D})}}{N[1+e^{8\pi\omega(M-\mathcal{D})}]}.
\end{eqnarray}
By tracing out $N-1$ particles, we obtain the reduced density operator of a single particle in the asymptotically flat region
\begin{eqnarray}\label{a7}
\rho_{1/2/\cdots/N-1}^{F,\mathrm{W}}=\frac{1}{N}\big(|1\rangle\langle1|+(N-1)|0\rangle\langle0|\big).
\end{eqnarray}
Following the definition of von Neumann entropy, we can calculate the entropies of density $\rho_{1/2/\cdots/N-1}^{F,\mathrm{W}}$, $\rho_{N_{\mathrm{out}}}^{F,\mathrm{W}}$ and $\rho_{123\ldots N_{\mathrm{out}}}^{F,\mathrm{W}}$, respectively. Substituting these into Eq.(\ref{a3}) then yields the mutual information for the dilaton-spacetime state $\rho_{123\ldots N_{\mathrm{out}}}^{F,\mathrm{W}}$ as
\begin{eqnarray}\label{S60}
I^{\mathrm{W}}_{F}(N)&=&-(N-1)(\frac{1}{N}\log_{2}\frac{1}{N}+\frac{N-1}{N}\log_{2}\frac{N-1}{N})-\frac{(N-1)e^{8\pi\omega(M-\mathcal{D})}}{N[1+e^{8\pi\omega(M-\mathcal{D})}]}\notag\\
&\times&\log_{2}\bigg\{\frac{(N-1)e^{8\pi\omega(M-\mathcal{D})}}{N[1+e^{8\pi\omega(M-\mathcal{D})}]}\bigg\}-\frac{N+e^{8\pi\omega(M-\mathcal{D})}}{N[1+e^{8\pi\omega(M-\mathcal{D})}]}\log_{2}\bigg\{\frac{N+e^{8\pi\omega(M-\mathcal{D})}}{N[1+e^{8\pi\omega(M-\mathcal{D})}]}\bigg\}\notag\\
&+&\frac{N-1}{N[1+e^{8\pi\omega(M-\mathcal{D})}]}\log_{2}\bigg\{\frac{N-1}{N[1+e^{8\pi\omega(M-\mathcal{D})}]}\bigg\}+\frac{1+Ne^{8\pi\omega(M-\mathcal{D})}}{N[1+e^{8\pi\omega(M-\mathcal{D})}]}\notag\\
&\times&\log_{2}\bigg\{\frac{1+Ne^{8\pi\omega(M-\mathcal{D})}}{N[1+e^{8\pi\omega(M-\mathcal{D})}]}\bigg\}.
\end{eqnarray}

\begin{figure}[t]
\centering
\begin{minipage}{0.48\linewidth}
\centering
\includegraphics[width=\linewidth]{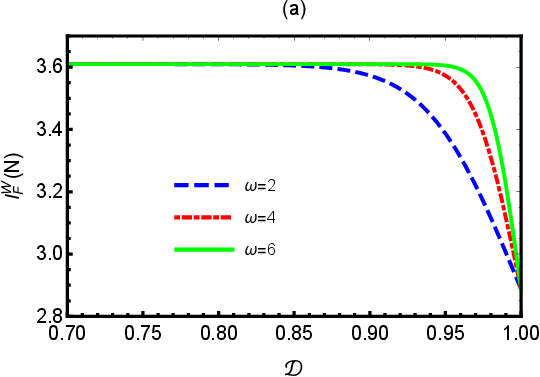}
\end{minipage}
\hfill
\begin{minipage}{0.48\linewidth}
\centering
\includegraphics[width=\linewidth]{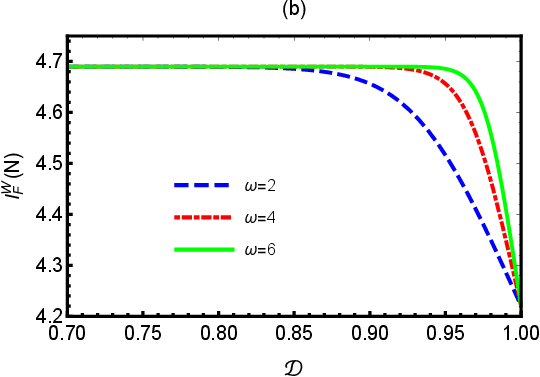}
\end{minipage}

\vspace{0.3cm}

\begin{minipage}{0.48\linewidth}
\centering
\includegraphics[width=\linewidth]{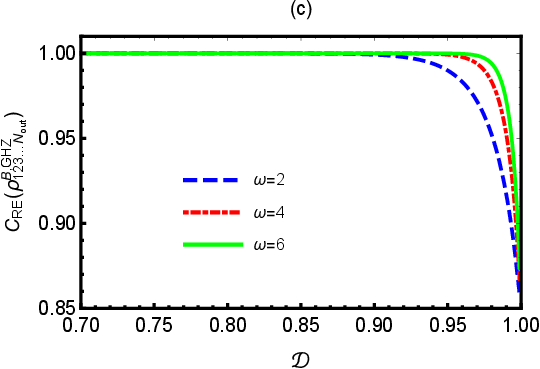}
\end{minipage}
\hfill
\begin{minipage}{0.48\linewidth}
\centering
\includegraphics[width=\linewidth]{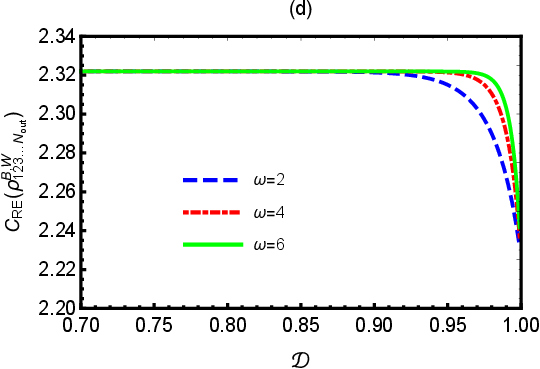}
\end{minipage}

\vspace{0.3cm}

\begin{minipage}{0.48\linewidth}
\centering
\includegraphics[width=\linewidth]{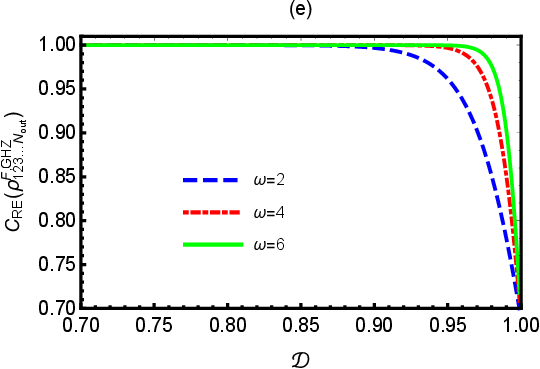}
\end{minipage}
\hfill
\begin{minipage}{0.48\linewidth}
\centering
\includegraphics[width=\linewidth]{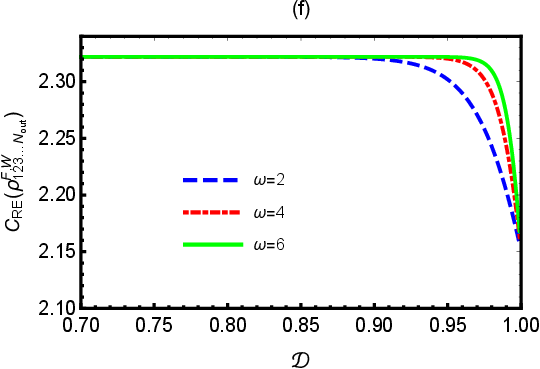}
\end{minipage}

\caption{Mutual information $I^{\mathrm{W}}_{F}(N)$ for fermionic W state with particle numbers $N=5$(a) and $N=10$(b), and REC for bosonic and fermionic GHZ and W states as a function of the dilaton parameter $\mathcal{D}$ for a fixed $N=5$ at different mode frequencies $\omega$.}
\label{Fig.4}
\end{figure}

In Fig.\ref{Fig.4}, we plot the mutual information $I^{\mathrm{W}}_{F}(N)$ for W states of the fermionic field and REC for GHZ and W states of the bosonic fermionic field as a function of the dilaton  $\mathcal{D}$.
It is important to note that the calculation of REC for the bosonic and fermionic fields of the GHZ and W states in the dilaton spacetime is detailed in Appendix A. Our analysis reveals a nontrivial interplay between particle statistics, multipartite state structure, and different  resources in curved spacetime. 
From Fig.\ref{Fig.1} to Fig.\ref{Fig.4}, we observe that fermionic systems retain more information than bosonic systems, indicating that the total correlations encoded in fermions are more resilient to gravitational degradation. However, their REC is smaller, reflecting a trade-off between total correlations and quantum coherence. Similarly, GHZ state consistently exhibits larger mutual information than W state, while their coherence is lower in curved spacetime. These results emphasize that both the particle statistics and the multipartite state structure play important roles in determining the behavior of different  resource measures in curved spacetime.

\begin{figure}[t]
\centering
\begin{minipage}{0.48\linewidth}
\centering
\includegraphics[width=\linewidth]{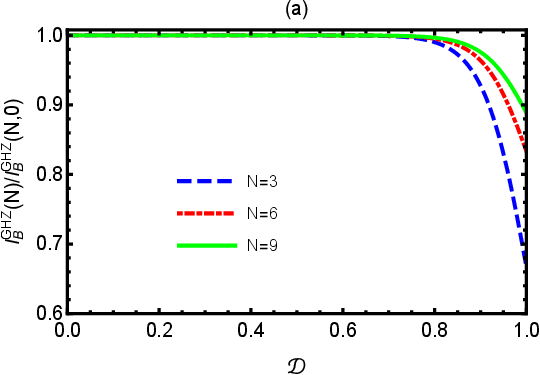}
\end{minipage}
\hfill
\begin{minipage}{0.48\linewidth}
\centering
\includegraphics[width=\linewidth]{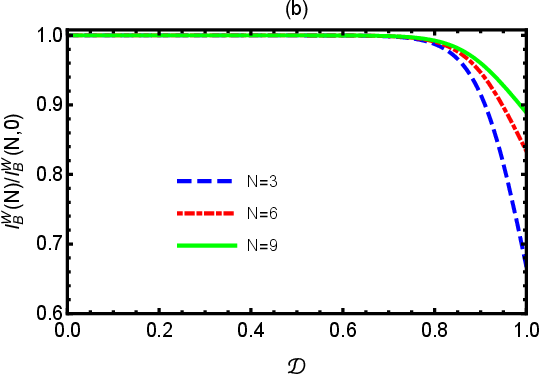}
\end{minipage}

\vspace{0.3cm}

\begin{minipage}{0.48\linewidth}
\centering
\includegraphics[width=\linewidth]{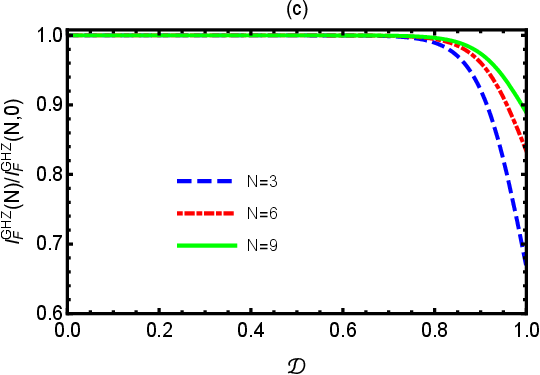}
\end{minipage}
\hfill
\begin{minipage}{0.48\linewidth}
\centering
\includegraphics[width=\linewidth]{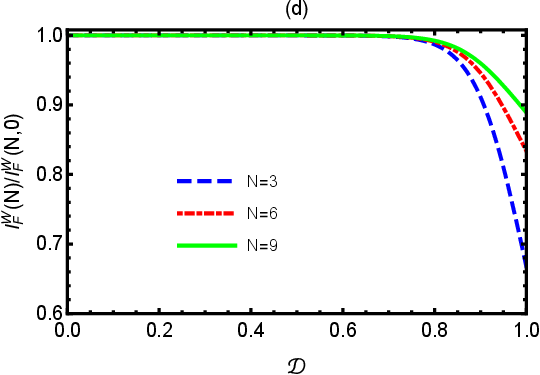}
\end{minipage}

\caption{The normalized multipartite mutual information $I(N)/I(N,0)$ as a function of the dilaton parameter $\mathcal{D}$ for a fixed $\omega=1$ at different particle numbers $N$.}
\label{Fig.5}
\end{figure}

To clarify the role of the particle number $N$ in multipartite mutual information, it is important to note that the quantity $I(\rho_{123\ldots N})$ defined in Eq.(\ref{a3}) naturally scales with the system size. 
Therefore, a direct comparison between systems with different particle numbers may not by itself reflect the robustness of correlations against the Hawking effect. To remove this trivial extensive contribution, we further analyze the normalized multipartite mutual information $I(N)/I(N,0)$.
 Our results show that, under the influence of the Hawking effect in the dilaton black hole background, the normalized mutual information $I(N)/I(N,0)$ increases gradually with increasing particle number $N$ in Fig.\ref{Fig.5}. This demonstrates that the average amount of total correlations carried by each particle becomes more robust in larger multipartite systems. In other words, although the Hawking effect continuously degrades  mutual information, larger systems exhibit a stronger resistance to dilaton-induced information loss, indicating an enhanced robustness of multipartite mutual information against gravitational effects.

\section{Conclusions}
In this work, we have systematically investigated the influence of the Hawking effect on 
the total amount of correlations of GHZ and W states by means of mutual information for both bosonic and fermionic fields in the background of the GHS dilaton black hole. We considered an operationally motivated scenario in which 
N observers initially share multipartite entangled states in the asymptotically flat region, after which one observer hovers near the event horizon while the remaining observers stay inertial at infinity. By tracing over the inaccessible field modes inside the event horizon, we derived general analytical expressions for the 
N-partite mutual information of both bosonic and fermionic fields in GHZ and W states within dilaton spacetime.

Our analysis reveals several essential features of multipartite  resources in strong gravitational fields by jointly examining multipartite mutual information. \textbf{(i) Particle statistics: mutual information vs REC:}  We find a clear contrast between bosonic and fermionic fields under the Hawking effect: fermionic systems retain larger multipartite mutual information than their bosonic counterparts, indicating a stronger robustness of total correlations against gravitational degradation, whereas their REC is consistently smaller than that of bosonic systems. For both statistics, the multipartite mutual information decreases monotonically with increasing dilaton, reflecting the gradual loss and redistribution of correlations induced by black hole radiation, and eventually saturates to a finite, mode-frequency-independent value in the extreme black hole limit, revealing a universal behavior dominated by spacetime geometry. \textbf{(ii) State structure: mutual information vs REC:} A similar complementary behavior is observed when comparing different multipartite state structures. GHZ state consistently exhibits larger multipartite mutual information than W state across the entire parameter range, while their REC remains smaller than that of W state. In addition, increasing the number of particles $N$ enhances the overall multipartite mutual information for both GHZ and W states in curved spacetime and improves its robustness against external gravitational effects.   \textbf{(iii) Particle number variation: multipartite mutual information vs entanglement:} Increasing the initial particle number enhances the multipartite mutual information of the GHZ and W states in the dilaton black hole background. However, the quantum entanglement of the GHZ state remains unaffected by the particle number, while the entanglement of the W state gradually decreases \cite{C10,C11}. This indicates that multipartite mutual information and entanglement respond differently to gravitational influences. Overall, these findings highlight a nontrivial complementarity between multipartite mutual information, quantum coherence, and entanglement in curved spacetime, demonstrating that different  resources react in qualitatively distinct ways to gravitational effects. Moreover, the robustness of total correlations does not necessarily imply the robustness of quantum coherence and entanglement.

Overall, our results demonstrate that both particle statistics and multipartite state structure play important roles in shaping the behavior of total correlations, entanglement, and quantum coherence in curved spacetime. These findings highlight the nontrivial interplay between different  correlation measures and gravitational effects in relativistic quantum systems. We expect that this work will contribute to a deeper understanding of many-body  correlation dynamics in strong gravitational fields and may provide useful insights for future studies of relativistic quantum information processing in curved spacetime.

\begin{acknowledgments}
 This work is supported by the National Natural Science Foundation of China (12575056) and  the Special Fund for Basic Scientific Research of Provincial Universities in Liaoning under grant NO. LS2024Q002.	
\end{acknowledgments}


\appendix
\onecolumngrid

\section{REC of GHZ and W states of bosonic and fermionic fields }
In this section, we calculate quantum coherence and obtain the corresponding analytical expressions in dilaton spacetime. To characterize coherence more comprehensively, we adopt the REC to quantify the distinguishability between a quantum state $\rho$ and its diagonal counterpart $\rho_{\text{diag}}$ in a given reference basis. The REC is defined as
\begin{eqnarray}\label{S71}
C_{RE}(\rho)=S(\rho_{\text{diag}})-S(\rho),
\end{eqnarray}
where $S(\rho)$ denotes the von Neumann entropy of state $\rho$, and $\rho_{\text{diag}}$ is obtained by removing all off-diagonal elements of $\rho$ in the reference basis. Using the above definitions, the REC corresponding to the states $\rho_{123\ldots N_{\mathrm{out}}}^{B,\mathrm{GHZ}}$ , $\rho_{123\ldots N_{\mathrm{out}}}^{B,\mathrm{W}}$, $\rho^{F,\mathrm{GHZ}}_{123\ldots N_{\mathrm{out}}}$, and $\rho_{123\ldots N_{\mathrm{out}}}^{F,\mathrm{W}}$ can be calculated as
\begin{eqnarray}\label{S72}
C_{RE}(\rho_{123\ldots N_{\mathrm{out}}}^{B,\mathrm{GHZ}})&=&-\frac{1}{2}\sum_{p=0}^{\infty}[1-e^{-8\pi \omega(M-\mathcal{D})}]e^{-8p\pi\omega(M-\mathcal{D})}\log_{2}\bigg\{\frac{1}{2}[1-e^{-8\pi \omega(M-\mathcal{D})}]\notag\\
&\times&e^{-8p\pi\omega(M-\mathcal{D})}\bigg\}-\frac{1}{2}\sum_{p=0}^{\infty}[1-e^{-8\pi \omega(M-\mathcal{D})}]^{2}(p+1)e^{-8p\pi\omega(M-\mathcal{D})}\notag\\
&\times&\log_{2}\bigg\{\frac{1}{2}[1-e^{-8\pi \omega(M-\mathcal{D})}]^{2}(p+1)e^{-8p\pi\omega(M-\mathcal{D})}\bigg\}+\frac{1}{2}\sum_{p=0}^{\infty}[1-e^{-8\pi \omega(M-\mathcal{D})}]\notag\\
&\times&e^{-8p\pi\omega(M-\mathcal{D})}\big[2+p-(p+1)e^{-8\pi\omega(M-D)}\big]\log_{2}\bigg\{\frac{1}{2}[1-e^{-8\pi \omega(M-\mathcal{D})}]\notag\\
&\times&e^{-8p\pi\omega(M-\mathcal{D})}\big[2+p-(p+1)e^{-8\pi\omega(M-\mathcal{D})}\big]\bigg\},
\end{eqnarray}

\begin{eqnarray}\label{S73}
C_{RE}(\rho_{123\ldots N_{\mathrm{out}}}^{B,\mathrm{W}})&=&-\frac{N-1}{N}\sum_{p=0}^{\infty}e^{-8p\pi\omega(M-\mathcal{D})}[1-e^{-8\pi\omega(M-\mathcal{D})}]\log_{2}\bigg\{\frac{1}{N}e^{-8p\pi\omega(M-\mathcal{D})}\notag\\
&\times&[1-e^{-8\pi\omega(M-\mathcal{D})}]\bigg\}-\frac{1}{N}\sum_{p=0}^{\infty}e^{-8p\pi\omega(M-\mathcal{D})}[1-e^{-8\pi\omega(M-\mathcal{D})}]^{2}(p+1)\log_{2}\bigg\{\frac{1}{N}\notag\\
&\times&e^{-8p\pi\omega(M-\mathcal{D})}[1-e^{-8\pi\omega(M-\mathcal{D})}]^{2}(p+1)\bigg\}+\frac{1}{N}\sum_{p=0}^{\infty}e^{-8p\pi\omega(M-\mathcal{D})}[1\notag\\
&-&e^{-8\pi\omega(M-\mathcal{D})}]\big[N+p-(p+1)e^{-8\pi\omega(M-\mathcal{D})}\big]\log_{2}\bigg\{\frac{1}{N}e^{-8p\pi\omega(M-\mathcal{D})}[1\notag\\
&-&e^{-8\pi\omega(M-\mathcal{D})}]\big[N+p-(p+1)e^{-8\pi\omega(M-\mathcal{D})}\big]\bigg\},
\end{eqnarray}

\begin{eqnarray}\label{S74}
C_{RE}(\rho^{F,\mathrm{GHZ}}_{123\ldots N_{\mathrm{out}}})&=&-\frac{1}{2}[e^{-8\pi\omega(M-\mathcal{D})}+1]^{-1}\log_{2}\bigg\{\frac{1}{2}[e^{-8\pi\omega(M-\mathcal{D})}+1]^{-1}\bigg\}-\frac{1}{2}[e^{8\pi\omega(M-\mathcal{D})}+1]^{-1}\notag\\
&\times&\log_{2}\bigg\{\frac{1}{2}[e^{8\pi\omega(M-\mathcal{D})}+1]^{-1}\bigg\}
-\frac{1}{2}\log_{2}\frac{1}{2}+\frac{1}{2[1+e^{8\pi\omega(M-\mathcal{D})}]}\notag\\
&\times&\log_{2}\bigg\{\frac{1}{2[1+e^{8\pi\omega(M-\mathcal{D})}]}\bigg\}+\frac{1+2e^{8\pi\omega(M-\mathcal{D})}}{2[1+e^{8\pi\omega(M-\mathcal{D})}]}\log_{2}\bigg\{\frac{1+2e^{8\pi\omega(M-\mathcal{D})}}{2[1+e^{8\pi\omega(M-\mathcal{D})}]}\bigg\},
\end{eqnarray}

\begin{eqnarray}\label{S75}
C_{RE}(\rho_{123\ldots N_{\mathrm{out}}}^{F,\mathrm{W}})&=&-\frac{N-1}{N}[e^{-8\pi\omega(M-\mathcal{D})}+1]^{-1}\log_{2}\bigg\{\frac{1}{N}[e^{-8\pi\omega(M-\mathcal{D})}+1]^{-1}\bigg\}-\frac{N-1}{N}\notag\\
&\times&[e^{8\pi\omega(M-\mathcal{D})}+1]\log_{2}\bigg\{\frac{1}{N}[e^{8\pi\omega(M-\mathcal{D})}+1]\bigg\}-\frac{1}{N}\log_{2}\big[\frac{1}{N}\big]\notag\\
&+&\frac{N-1}{N[1+e^{8\pi\omega(M-\mathcal{D})}]}\log_{2}\bigg\{\frac{N-1}{N[1+e^{8\pi\omega(M-\mathcal{D})}]}\bigg\}+\frac{1+Ne^{8\pi\omega(M-\mathcal{D})}}{N[1+e^{8\pi\omega(M-\mathcal{D})}]}\notag\\
&\times&\log_{2}\bigg\{\frac{1+Ne^{8\pi\omega(M-\mathcal{D})}}{N[1+e^{8\pi\omega(M-\mathcal{D})}]}\bigg\}.
\end{eqnarray}

\end{document}